_

\documentclass[preprint]{aastex63}

\usepackage{hyperref}
\usepackage{color,natbib}
\usepackage[T1]{fontenc}
\def\apjl{ApJ}%
\def\apjs{ApJS}%
\def\aap{A\&A}
\def\rgs{Radio Galaxies~}
\def\lg{LERG}
\def\lgs{LERGs}

\def\hgs{HERGs}
\def\g{$\eta/\epsilon$}

\def\FRcats{FR catalogs}
\usepackage[normalem]{ulem}

\usepackage{times,epsfig,natbib,amssymb,amsmath,graphicx,microtype}
\usepackage{array,multirow}
\usepackage{hyperref}
\usepackage{color,natbib}

\revised{December 22, 2020}
\accepted{February 15, 2021}
\shorttitle{Radio Galaxies in  the nearby mJy Universe}
\shortauthors{Grandi et al.}
\graphicspath{{./}{figures/}}

\begin{document}

\title{Jet-accretion system in the nearby mJy Radio Galaxies}

\correspondingauthor{Paola Grandi}
\email{paola.grandi@inaf.it}
\author[0000-0003-1848-6013]{Paola Grandi}
\affiliation{INAF-Osservatorio di Astrofisica e Fisica dello Spazio, Via Gobetti 101, I-40129, Bologna, Italy}
\author[0000-0002-5201-010X]{Eleonora Torresi}
\affiliation{INAF-Osservatorio di Astrofisica e Fisica dello Spazio, Via Gobetti 101, I-40129, Bologna, Italy}
\author{Duccio Macconi}
\affiliation{Dipartimento di Fisica e Astronomia, Universit\'a degli Studi di Bologna, via Gobetti 93/2, I-40129 Bologna, Italy}
\affiliation{INAF-Osservatorio di Astrofisica e Fisica dello Spazio, Via Gobetti 101, I-40129, Bologna, Italy}
\author{Bia Boccardi}
\affiliation{Max-Planck-Institut f\"{u}r Radioastronomie, Auf dem H\"{u}gel 69, D-53121 Bonn, Germany}
\author{Alessandro Capetti}
\affiliation{INAF-Osservatorio Astrofisico di Torino, Strada Osservatorio 20, I-10025, Pino Torinese, Italy}



\begin{abstract}
It is generally thought that FRII \rgs host thin optically thick disks, while FRIs are powered by Advected Dominated Accretion Flows. 
The sources with an efficient engine are optically classified as High Excitation Radio Galaxies (\hgs) and those with an inefficient motor as Low Excitation Radio Galaxies (\lgs).
Recently, the study of \rgs down to mJy fluxes has cast serious doubts on the LERG-FRI and HERG-FRII correspondence, revealing that many LERGs show  FRII radio morphologies.\\
The FR catalogs recently compiled by \cite{FRIcat, FRIIcat} and \cite{FR0cat} have allowed us to explore this issue in the local ($z\le 0.15$) mJy Universe. Our statistical study shows that the majority of nearby mJy objects are in a late stage of their life.  FRII-LERGs appear more similar to the old FRI-LERGs than to the young FRII-HERGs.
FRII-LERGs may be aged HERGs that, exhausted the cold fuel, have changed their accretion regime or a separate LERG class particularly efficient in launching jets. 
Exploiting the empirical relations which convert  L$_{\rm [OIII]}$ and L$_{\rm 1.4~GHz}$  into accretion power and jet kinetic power, respectively, we observed that LERGs with similar masses and accretion rates seem to expel jets of different power.  We speculate that intrinsic differences related to the black hole properties (spin and magnetic field at its horizon) can determine the observed spread in jet luminosity. In this view,  FRII-LERGs should have the fastest spinning black holes and/or the most intense magnetic fluxes. On the contrary, compact LERGs (i.e. FR0s)  should host
extremely slow black holes and/or weak magnetic fields.

\end{abstract}


\section{Introduction}\label{sec:intro}
Radio Galaxies (RGs) are historically divided in core-brightened FR~I and bright edge-brightened FR~II \citep{fanaroff74}, on the basis of their extended radio morphology that approximately changes at a critical power P$_{\rm 1.4~ GHz} \sim3$ $\times$ 10$^{25}$ W~Hz$^{-1}$. \cite{led96} refined the classification showing that the break between FRIs and FRIIs is a strong function of the  host galaxy absolute magnitude ($M_{R}$). As the host galaxy luminosity traces the black hole mass \citep{mag98} and the radio power is proportional to the accretion luminosity \citep{willott99}, the FRI-FRII separation was later re-interpreted in terms of accretion rates \citep{ghisellini01}. The less powerful radio galaxies (i.e. FRIs) host an inefficient hot thick flow, while the more powerful sources  (i.e. FRIIs)   have an efficiently accreting cold disk. In support of this interpretation, \cite{marchesini04} found an accretion rate gap between FRIs and FRIIs, suggestive of a different accretion regime. 
From the optical point of view, radio galaxies are split into High Excitation Radio Galaxies (HERGs) and Low Excitation Radio Galaxies (LERGs) \citep{JaR}, with LERGs characterized by [OIII] equivalent width $<10$~\AA~ and/or [OII]/[OIII] ratios $>1$. More recently, \cite{buttiglione10} proposed  a combination of emission lines, the excitation index (EI\footnote{$EI=Log([OIII]/H_{\beta})-1/3(Log([NII]/H_{\alpha})+ Log([SII]/H_{\alpha}) +Log([OI]/H_{\alpha}))$}), to distinguish the classes:
LERGs have  $EI<0.95$ and  HERGs $EI>0.95$.  \\
As FRIIs are generally associated to HERGs and  FRIs to LERGs, it is almost natural to consider the nuclear engine as the main driver of the FRI-FRII dichotomy.
However, this one-to-one correspondence (FRI-LERGs versus FRII-HERGs), based on the study of powerful sources with Jy flux densities, is probably a simplification.

For example,  24 FRII sources in the 3CR sample \citep{buttiglione10} lack high excitation emission lines and are classified as LERGs. Similarly, \cite{tad98} studying the 2Jy sample \citep{wp1985} found that 23\% of the FRIIs  are  Weak Line Radio Galaxies (WLRGs), i.e. objects with EW$_{\rm [OIII]} < 10$~\AA.  As discussed by \cite{Tad16}, WLRGs generally correspond to  LERGs, although the classification criteria are slightly different. 
Moreover, some FRIs have efficient accretion disks (i.e. they are FRI-HERGs). 3C~120, with broad and intense optical lines, a prominent UV bump, and an iron line in the X-ray spectrum \citep{torresi12a}, is a typical example.
The difficulty in reconciling accretion mode and kpc radio morphology has become more evident in recent years when large-area radio (NVSS, FIRST) and optical (SDSS and 6dFGRS) spectroscopic surveys have allowed expanding the study of radio galaxies down to mJy fluxes (see \cite{heckman14} for a review). Several studies show that radio galaxies with FRII morphologies preferentially host low efficient accretion flows (i.e. they are classified as LERGs) at low flux densities \citep{FRIcat, FRIIcat, miraghaei}. Finally, a recent analysis of low luminosity radio galaxies observed by LOFAR \cite{Mingo} has questioned the FRI/FRII break based on the radio power. At low fluxes, any association between morphology and radio luminosity seems to disappear.
If radio galaxies of similar radio morphology (radio power) can come into different "accretion flavors", new scenarios have to be considered.
The accretion rate could not be the driving parameter and something else related to the black hole could play a major role in launching the jet \citep{ghisellini14}. 
The environment could be also  important, as radio, optical and  X-ray  studies \citep{cro2005,Gawronski,cro2008, gendre,ineson, cro2018,Mingo,duccio} seem to suggest.
Finally, we could be observing  different phases that AGN pass through their life.
For example, a recent X-ray analysis of 3C radio galaxies \citep{duccio} has shown that FRII-LERG nuclei have less cold gas, i.e. smaller column densities (N$_{\rm H}$) than FRII-HERGs. A possible suggestion is that a transition occurs from a thin disk to a thick flow in FRIIs when the cold fuel has been depleted. 
Incidentally, this leads to speculate that FRIs could switch from LERG to HERG if a sudden replenishment of fresh cold gas occurs, maybe due to a galaxy merger (see, for example, \cite{garofalo19}).
However, most of the results (and controversial interpretations) are based on the study of very bright (Jy) radio sources which make up only a small fraction of the total radio galaxy population.
In order to shed light on these open questions, the jet-accretion system is explored through the study of local faint (mJy) radio galaxies taking advantage of the recently compiled FR catalogs {FRcat} \citep{FRIcat, FRIIcat, FR0cat} that include sources well characterized both in the radio and optical bands.\\
A cosmology with $H_0 = 67$ km s$^{-1}$ Mpc$^{-1}$, $\Omega_m = 0.32$, and $\Omega_\Lambda = 0.68$ \citep{planck} is assumed in this paper.

\section{The FRcat samples}

The FR0 \cite{FR0cat}, FRI \citep{FRIcat}   and FRII \citep{FRIIcat}  catalogs include 108, 219 and 122 radio galaxies, respectively.
They are part of a large sample  assembled by \cite{best12} (hereafter B12 sample),  cross-correlating
the seventh data release of the Sloan Digital Sky Survey (SDSS) with the NRAO (National
Radio Astronomy Observatory) VLA (Very Large Array) Sky Survey (NVSS) and the Faint
Images of the Radio Sky at Twenty centimeters (FIRST) survey.  For all the AGN in the sample, \cite{best12} provided an optical classification (LERG/HERG)  assuming different criteria \citep{buttiglione10,kewley,cid} depending on the signal to noise ratio of emission lines (e.g. H$_{\alpha}$, H$_{\beta}$, OIII, OI,  NII, SII)  revealed in the SDSS spectra.

The FRI and FRII catalogs are limited to local radio galaxies (maximum distance z=0.15) with an NVSS flux density larger than 5~mJy and a (one-side) extension of at least 30~kpc\footnote{1" corresponds to 2.72 kpc at z=0.15.}. The radio classification was  performed by a visual inspection of the FIRST images.
 If a radio galaxy showed a higher surface brightness near the core (edge-darkness), it was defined as FR Type I. On the contrary, if it appeared brighter at the end (edge-brightened), the associated radio class was FR Type II. 
An additional sample of 14 small FRIs (sFR) was also included in the  FRIcat.  It consists of sources located at z$\le 0.05$  and with a radio extension between 10 and 30~kpc.
As stressed by \cite{FRIIcat}, the FRI and FRII catalogs are statistically complete at a level of $\sim90\%$ in the optical range and have a flux limit of $\sim 50$~mJy at 1.4~GHz.

The FR0 catalog \citep{FR0cat}  consists of FIRST compact radio galaxies with a minimum flux density of 5~mJy at 1.4 GHz, at redshift $\le0.05$ (i.e. with a maximal radio extension of 2.5~kpc), all optically classified as LERGs (see \cite{baldigalaxy} for a review on this class of objects). Four compact sources with HERG properties were also revealed but not included. 
A summary of the selection criteria is reported in Table 1.

\begin{figure}
 \plotone{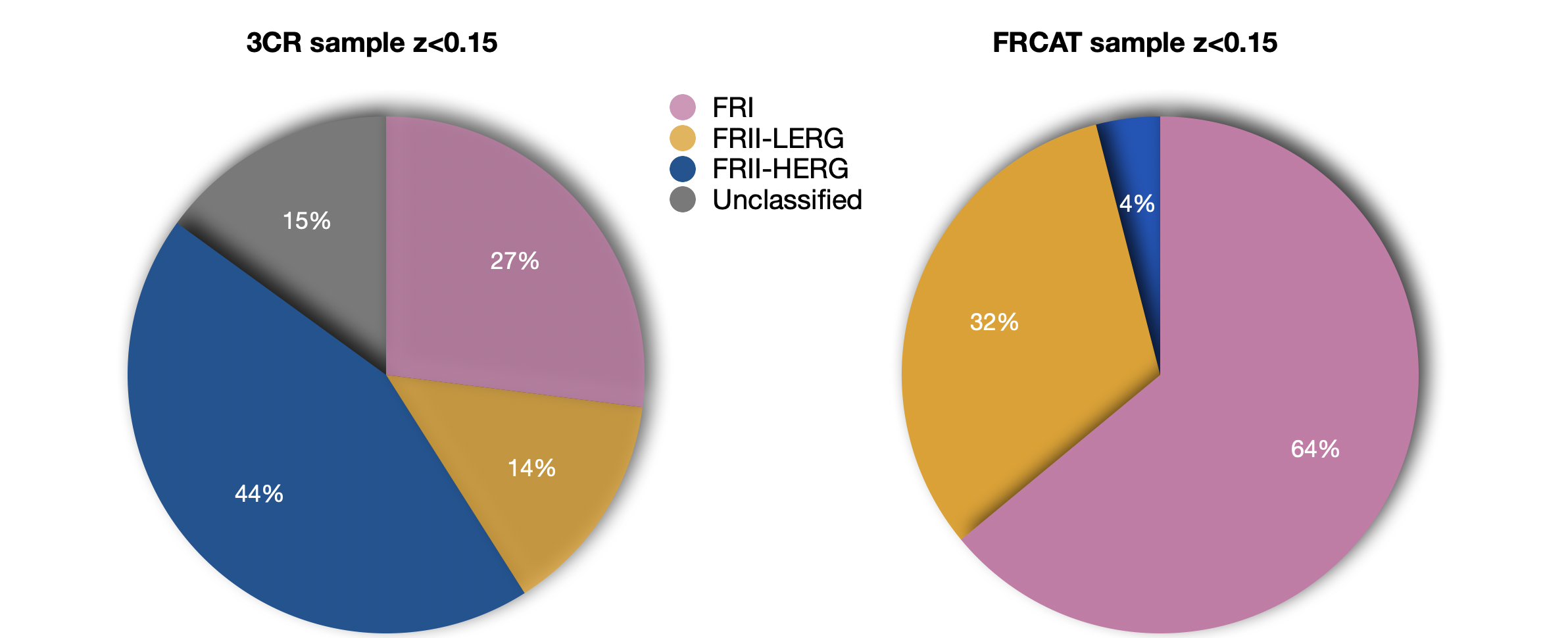}
\caption{Fraction of FRI and FRII radio galaxies with  z$<0.15$ in the 3CR ({\it left pie}) and in the FRcat ({\it right pie})
catalogs. The number of LERGs  increases going down to mJy fluxes.}
\label{population}
\end{figure}

\begin{table}
\begin{center}
\caption{Sample selection criteria: $F_{\it 1.4\rm GHz}>5$mJy}
\begin{tabular}{llll}
\hline
\hline 
Sample & z &\multicolumn{1}{l}{Optical} &\multicolumn{1}{c}{Extension}  \tabularnewline
       & & \multicolumn{1}{l}{Class}& \multicolumn{1}{c}{(kpc)} \tabularnewline
 \hline
FR0cat              & $<0.05$ & LERG& $< 2.5$ \tabularnewline
 FRIcat             & $<0.15$ &LERG &$>30$  \tabularnewline
 sFRIcat             & $<0.05$ &LERG &$>10$ and $<30$  \tabularnewline
 FRIIcat              &$<0.15$ & LERG/HERG &$>30$  \tabularnewline
\hline\hline
\end{tabular}
\end{center}
\label{criteria}
\end{table}

While in the 3CR catalog more than 40\% are powerful radio galaxies with an efficient accretion disk, in the FR catalogs  \rgs with high excitation emission lines are a minority $\sim 4\%$ (see Figure 1).
Interestingly enough, \cite{miraghaei} provided the radio classification of another  BH12 subsample adopting slightly different selection criteria.
Despite the different approaches, they confirm the predominance of LERGs in the FRII population.

\subsection{SDSS observables and  derivated quantities}
Thanks to the MPA-JHU DR7 release of spectrum  measurements\footnote{\url{https://wwwmpa.mpa-garching.mpg.de/SDSS/DR7/}}, we could estimate black hole mass  and radiative luminosity for each source of the \FRcats.
The velocity dispersion ($\sigma_{\ast}$) was converted into BH mass using the relation $Log(M_{BH}/M_{\odot})=8.32+5.64 \times Log(\sigma_{\ast}/200~$km~s$^{-1})$ \citep{mcconnell} and the [OIII]$\lambda$5007 luminosity into radiative luminosity (hereafter named accretion luminosity, $L_{\rm acc}$) using the multiplicative factor 3500, L$_{\rm acc}=3500\times L_{\rm [OIII]}$ \citep{Heckman04}.

Other important DR7 quantities, useful to characterize  the galaxies hosting different FR classes, are the stellar mass
and the calcium break D$_n$4000.
The stellar masses are obtained fitting a large grid of models from \cite{stefan} to the broad-band u,g,r,i,z photometry of SDSS. 
The calcium break values are derived considering the ratio of  the flux  in the red continuum (4000-4100 \AA)  to that in the blue continuum (3850-3950 \AA) \citep{balog}.  The  D$_n$4000 jump is considered an indicator of the stellar activity  \citep{worthey}.  Being due to metal absorption, it is expected to be smaller in galaxies with young stars (i.e. with highly ionised atoms).

\subsection{Checking the FR0 sample}
The selection and classification of large samples of objects necessarily imply the inclusion of a small fraction of spurious sources. 
As noted by \cite{best12}, this is not a problem if the number of sources is large (of the order of several hundreds or more), but it could have an impact on smaller samples. 
This is particularly true for the FR0s that are not resolved in the FIRST survey and could be misclassified.
Low radio flux density sources without any resolved jet structure could hide a weak BL Lac nucleus or a radio-quiet LINER with intense star-formation.

\begin{figure}[ht!]
\plottwo{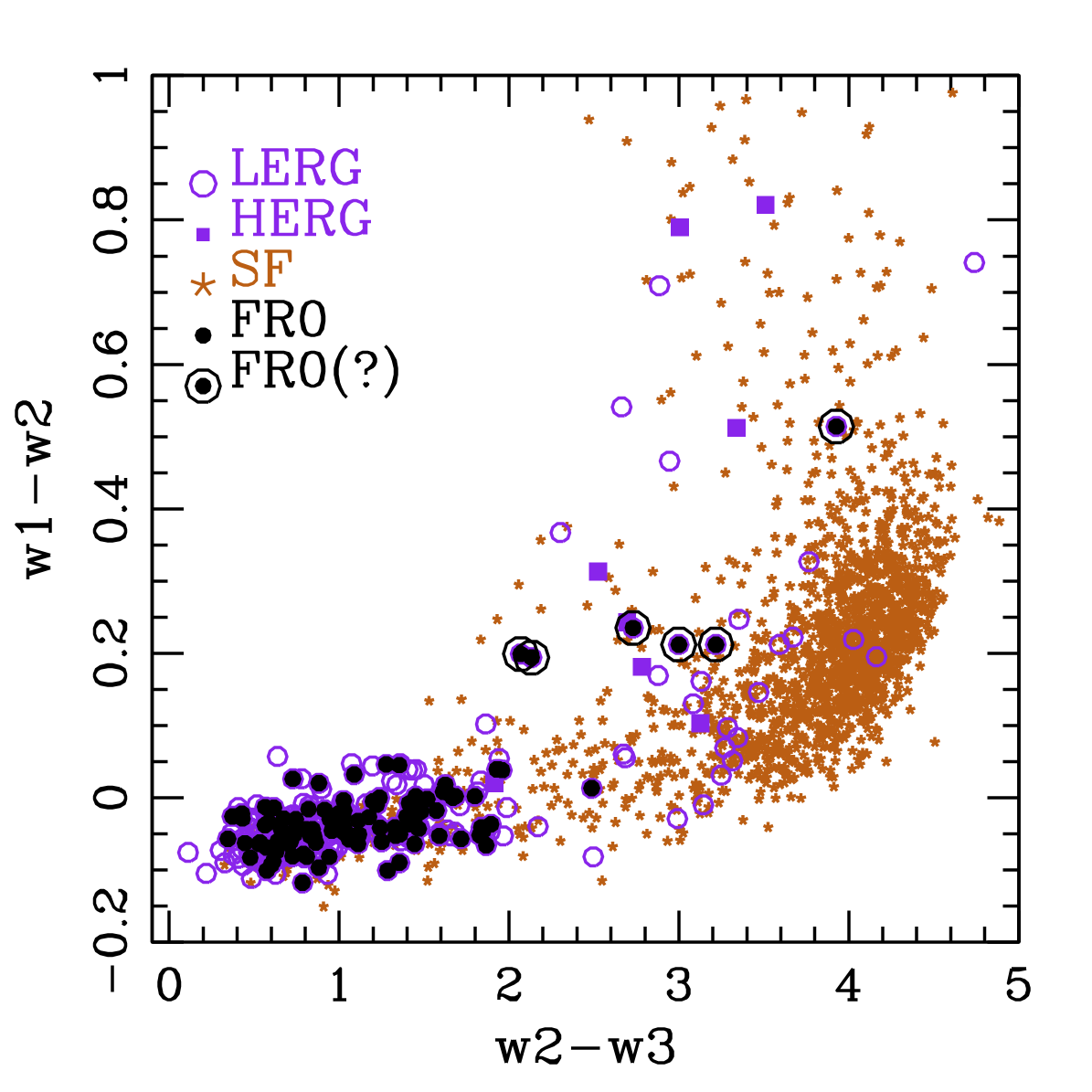}{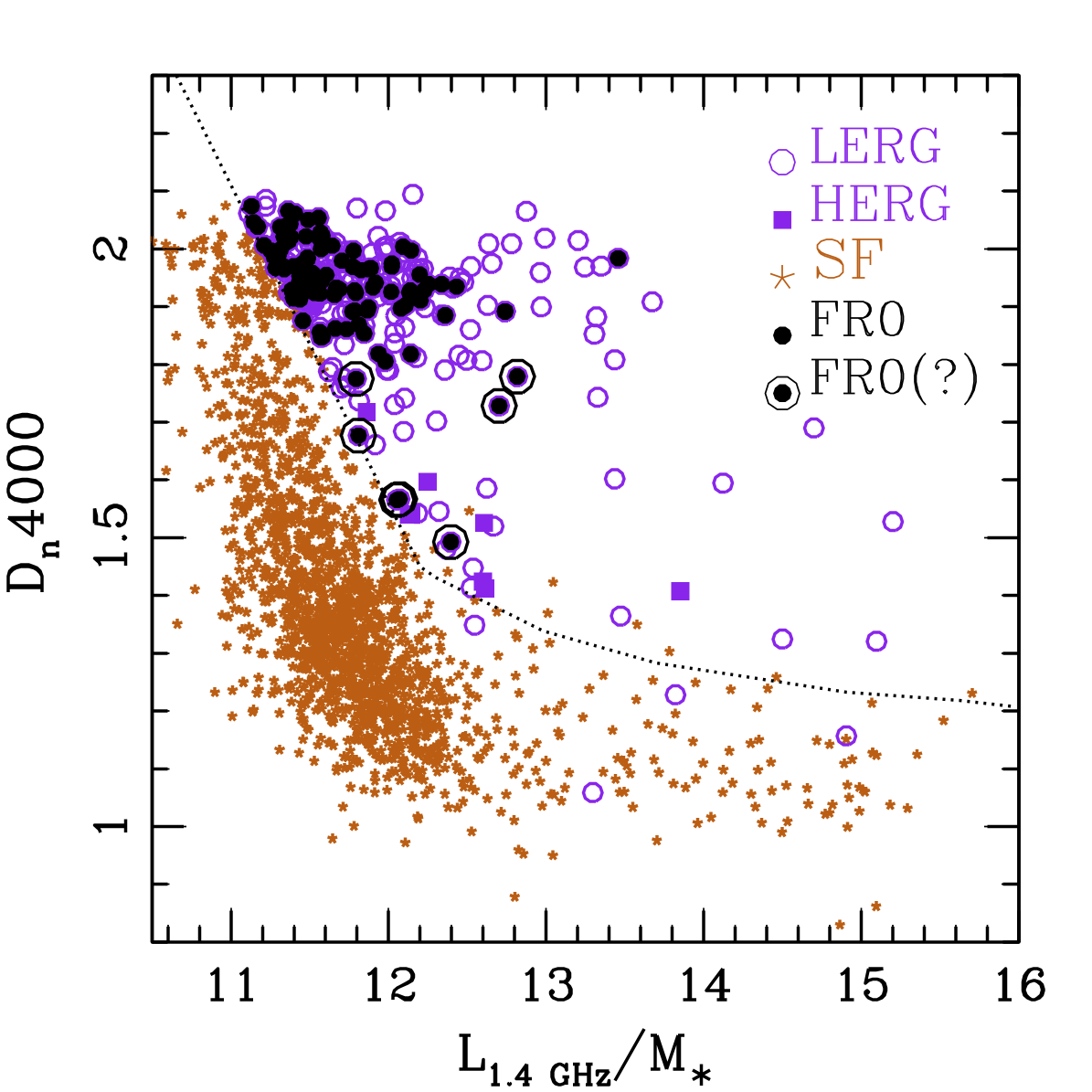}
\caption{Diagnostic diagrams of radio sources with z$\le 0.05$ of the sample of \cite{best12}. LERGs are purple open circles, HERGs purple squares and Star-forming galaxies (SF) orange stars. FR0s from \cite{FR0cat}  are marked by black points.
{\it Left panel} -- WISE color-color diagram.  LERGs are in the lower right corner, while the few HERGs are shifted towards redder colours. SF on the left are spread on a wide zone. FR0s mainly fall in the LERG zone.
Those with w3-w2$>2$ and w2-w1$>0.1$ (open black squares) are excluded by the sample as possible spurious objects.
{\it Right panel} -- D$_n$4000 versus $L_{1.4GHz}/M_{\ast}$ diagnostic plane.  The majority of FR0s are clustered in the AGN region with D$_n$4000$\sim 2$. The sources with D$_n$4000$<1.8$ are rejected as possible Star-forming galaxies or BL Lacs. }.

\label{clean}
\end{figure}

We sought for possible spurious sources exploring the WISE color-color diagram.
Three infrared bands, w1(3.4$\mu$m), w2(4.6$\mu$ m), and w3(12$\mu$ m) were considered. The color w3-w2 was plotted versus the color w2-w1. It is known that different sources occupy different regions of the plot with redder objects characterized by higher values of w3-w2 and w1-w2. Elliptical galaxies are expected to have colors near zero while star-forming galaxies (SF) are very red in both w3-w2 and w2-w1. Radio quiet and radio loud AGN  with efficient accretion disks and dusty screens are in between (see Fig. 12 of \cite{2010w}).

Figure \ref{clean} ({\it left panel}) shows all the objects of the B12 sample with z$\le 0.05$ together with FR0s marked as black points. LERGs are in the elliptical region, star-forming galaxies mainly above w3-w2 $>2$ and the few HERGs in the AGN area.
As expected there is overlapping between FR0s and LERGs, although a handful of compact radio sources are shifted to redder colors (w3-w2$>2$ and w2-w1$>0.1$).
Some of them are also clearly separated from the bulk of the FR0 population in the D$_n$4000 versus $L_{\rm 1.4GHz}/M_{\ast}$ diagram (Fig. \ref{clean}--{\it right panel}), one of the diagnostic plots proposed by \cite{best12} to divide AGN from star-forming galaxies.

Taking a conservative approach, we then decided to exclude from our statistical analysis those FR0s with redder WISE colours (w3-w2$>2$ and w2-w1$>0.1$). These have a not negligible probability to be star forming galaxies.
Also, we also excluded FR0s with D$_n$4000 $<1.8$.
As pointed out by \cite{cr15}, a small amplitude of the $4000$\AA~ break could be a signature not only of young stars but also of a jet. The non-thermal radiation can indeed dilute the optical continuum reducing the D$_n$4000 depth.
Finally, we note that the not genuine FR0 nature is certain for, at least, two compact radio galaxies. They have red WISE colours, small calcium break, a BH mass less than $10^8$ M$_{\odot}$ (typical of radio-quiet AGN), and a high probability to reside in spiral galaxies \citep{H-C2011,tempel}.

At the end of this selection, the "clean" sample consisted of 99 FR0s with only 9 rejected sources (less than 1\% of the sample).

\section{Comparison among the different classes}

\begin{figure}[bh!]

\plottwo{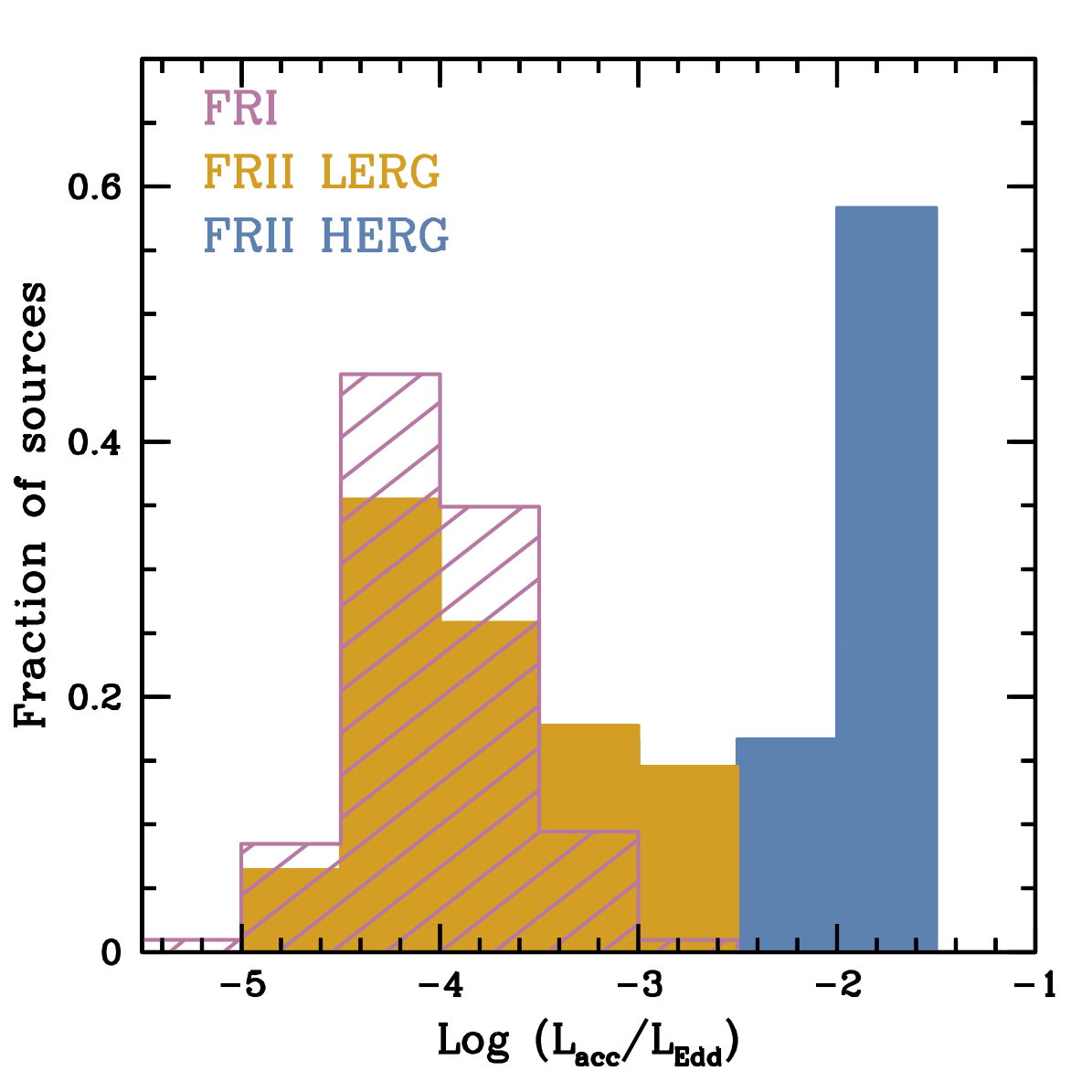}{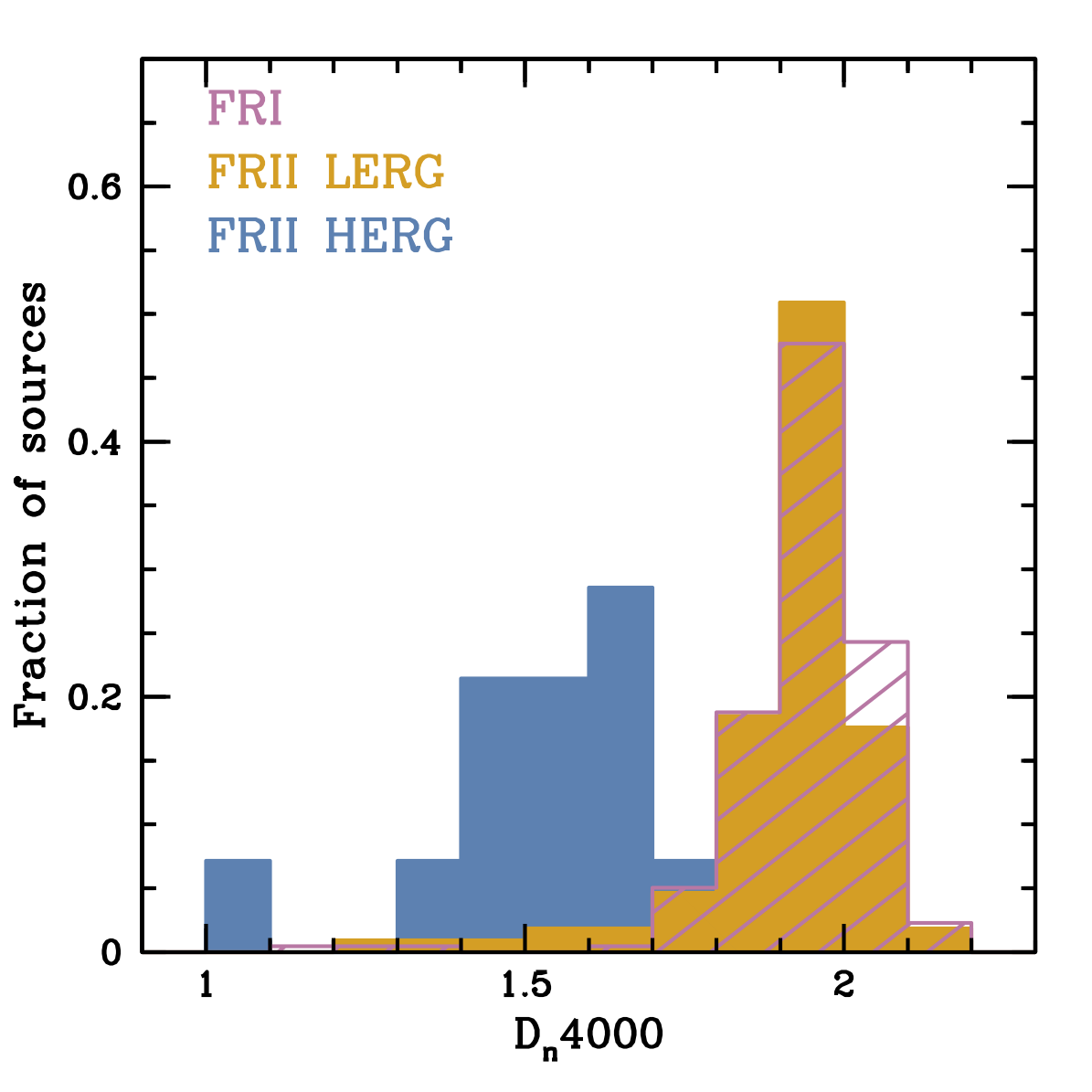}
\caption{Eddington-scaled radiative luminosity ({\it left panel}) and 
calcium break ({\it right panel}) histograms. FRII-LERGs are similar to FRIs. Both live in old galaxies and are powered by hot accretion flows. FRII-HERGs  are more radiatively efficient and show signs of star-forming activity.}
\label{histo}
\end{figure}

Table \ref{stat} reports median, average, and relative standard deviation of all the studied quantities. A comparison among the different classes was performed by applying a Kolmogorov-Smirnov (KS) test. We conservatively assumed that two data sets are different if the Kolmogorov-Smirnov  probability is less than $\times 10^{-3}$. In other words, we reject the null hypothesis that the two data sets are drawn from the same distribution at a confidence level $> 3\sigma$. The KS results are in Table \ref{ks}. In Figure \ref{histo} the most interesting histograms are shown.

\begin{table}
\centering
\caption{FRCat Average Properties} 
\begin{tabular}{|l|lllc|l|lllc|}
\hline
\hline
\textcolor{black}{\bf Class} &\textcolor{black}{\bf median} &\textcolor{black}{\bf average} &\textcolor{black}{\bf std} &  \textcolor{black}{\bf N. objects} &\textcolor{black}{\bf Class}&\textcolor{black}{\bf median} & \textcolor{black}{\bf average} & \textcolor{black}{\bf std} &  \textcolor{black}{\bf N.objects} \\
\hline
 &&&&&&&&&\\
\textcolor{black}{($z<0.15$)}&\multicolumn{4}{c|}{ Log (L$_{1.4GHz}$) (erg s$^{-1}$) }&\textcolor{blue}{($z<0.05$)}&\multicolumn{4}{c|}{ Log($L_{1.4GHz}$) (erg s$^{-1}$)}\\
\hline
\textcolor{black}{FRI} & 40.32 & 40.34 & 0.34& 219  &\textcolor{blue}{FR0} & 38.87 & 38.96 &0.36 & 99\\
\textcolor{black}{FRII-LERG} & 40.77 & 40.75 & 0.49& 108 &\textcolor{blue}{FRI}& 40.12&  40.13&  0.40 &  9\\ 
\textcolor{black}{FRII-HERG} & 41.37 & 41.26& 0.55& 14 & \textcolor{blue}{small FRI}& 39.52&  39.60&  0.34  &  14\\
 \hline
  &&&&&&&&&\\
 &\multicolumn{4}{c|}{Log($L_{\it OIII}$) (erg s$^{-1}$)}& &\multicolumn{4}{c|}{Log($L_{\it OIII}$) (erg s$^{-1}$)} \\
\hline
\textcolor{black}{FRI} & 39.87 &  39.86 & 0.23& 107 &\textcolor{blue}{FR0} & 39.60& 39.58& 0.27  &    98 \\
\textcolor{black}{FRII-LERG} & 39.91 &  39.95&  0.39& 66& \textcolor{blue}{FRI} & 39.51& 39.50& 0.17  &    9 \\
\textcolor{black}{FRII-HERG} & 41.42 & 41.30 & 0.52& 14 & \textcolor{blue}{small FRI}&39.39   & 39.37 &0.15 &12\\
\hline
&&&&&&&&&\\
&\multicolumn{4}{c|}{Velocity Dispersion $\sigma_{\ast}$ (km s$^{-1}$)}&&\multicolumn{4}{c|}{Velocity Dispersion $\sigma_{\ast}$   (km s$^{-1}$)}\\
\hline
\textcolor{black}{FRI} &     254  &    256 &     35 &      218 &\textcolor{blue}{FR0} &  231  &    237 &  38 &     97 \\
\textcolor{black}{FRII-LERG} &   246  &    243 &      39   &   102 & \textcolor{blue}{FRI}&  270&    276&      31 &     9\\
\textcolor{black}{FRII-HERG} &  209 &   204 &       23   &      12&\textcolor{blue}{small FRI} &  253  &    261 &      32  &     14 \\
 \hline
&&&&&&&&&\\
 &\multicolumn{4}{c|}{Stellar Mass Log($M_{\ast}$) (M$_{\odot}$)}&& \multicolumn{4}{c|}{Stellar Mass Log($M_{\ast}$) (M$_{\odot}$)}\\
\hline
\textcolor{black}{FRI} &    11.38  &  11.38 &  0.19  &      214& \textcolor{blue}{FR0} &11.15    & 11.13  & 0.22 &   97 \\
\textcolor{black}{FRII-LERG} &    11.33  &  11.33 &  0.31  &     107 &\textcolor{blue}{FRI} &11.32   & 11.34 & 0.13 &   9\\
\textcolor{black}{FRII-HERG} &    11.09  &  11.08 &  0.25 &     14 &\textcolor{blue}{small FRI}  & 11.22  & 11.23  & 0.17 & 14  \\
\hline
&&&&&&&&&\\
&\multicolumn{4}{c|}{Calcium break (D$_n$4000)} &&\multicolumn{4}{c|}{Calcium break (D$_n$4000)}\\
\hline
\textcolor{black}{FRI} &  1.96 &1.94&     0.12&         218&\textcolor{blue}{FR0}&  1.96 &1.95&     0.06&        99\\
\textcolor{black}{FRII-LERG} &  1.96 &1.94&     0.14&        108 &\textcolor{blue}{FRI} &  1.95 &1.91&     0.07&        9\\
\textcolor{black}{FRII-HERG} &  1.59 &1.53&     0.19&        14& \textcolor{blue}{small FRI}&  1.96 &1.94&     0.09&        14\\
\hline
&&&&&&&&&\\
&\multicolumn{4}{c|}{Log(BH) (M$_{\odot}$)}&&\multicolumn{4}{c|}{Log(BH) (M$_{\odot}$)}\tabularnewline
\hline
\textcolor{black}{FRI} &  8.91 & 8.91&     0.39&         218 &\textcolor{blue}{FR0} &  8.67 & 8.71&     0.39&         97 \\
\textcolor{black}{FRII-LERG} &  8.83 & 8.77&     0.41&         102&\textcolor{blue}{FRI} &  9.06& 9.10&     0.27&         9\\
\textcolor{black}{FRII-HERG} &  8.43& 8.36&     0.28&         12&\textcolor{blue}{small FRI} &  8.89& 8.95&     0.29&         14 \\
\hline
&&&&&&&&&\\
&\multicolumn{4}{c|}{Log($L_{acc}/L_{Edd}$)} &&\multicolumn{4}{c|}{Log($L_{acc}/L_{Edd}$)}\\
\hline
\textcolor{black}{FRI} &  -4.03 & -4.03& 0.40& 106 &\textcolor{blue}{FR0}&  -3.99& -4.04 &0.47& 96\\
\textcolor{black}{FRII-LERG} &  -3.85& -3.74& 0.55 & 62&\textcolor{blue}{FRI} &  -4.38 & -4.50& 0.32& 12 \\
\textcolor{black}{FRII-HERG} &  -1.84& -2.1 & 0.57& 12 &\textcolor{blue}{small FRI}&  -4.38 & -4.50& 0.32& 12 \\
\hline
\hline
\end{tabular}
\label{stat}
\end{table}
\noindent

As expected, FRII-HERGs and FRIs are distinct populations. 
FRII-HERGs have smaller black holes, a larger accretion rate expressed in terms of $L_{\rm acc}/L_{\rm Edd}$  and more stellar activity. They are younger systems.  More interesting is that our analysis shows that FRII-HERGs and FRII-LERGs are also different.
FRIIs with an inefficient engine have more massive black holes and a more evolved stellar population (Table \ref{stat} and Fig. \ref{histo}), i.e. are more similar to FRIs. Indeed, FRI and FRII-LERG classes are almost completely overlapped in the histograms of Fig. \ref{histo}.\\
The nuclear properties of mJy LERG sources, independently of their radio morphology, are very similar, at odds with the trend observed in Jy radio galaxies.
As shown by  \cite{duccio}, the FRII-LERGs of the 3C sample have indeed accretion rates ($L_{\rm acc}/L_{\rm Edd}$) generally lower than FRII-HERGs but higher than FRIs. 

At $z\le 0.05$ no significant difference is observed between small and extended FRIs. 
FR0s have accretion rates slightly higher than FRIs (see Tab \ref{stat}). Although potentially interesting, we do not further speculate on this result, as still missed outliers in the FR0 sample can not be definitively excluded.  However, recent stellar activity is not observed in any classes, suggesting that all the LERGs in the Local Universe are in a late stage of evolution.
\begin{table*}
\centering
\caption{FRCat Kolmogorov Smirnov test results} 
\begin{tabular}{|l|llllll|}

\hline 

\multicolumn{1}{|l}{\textcolor{black}{\bf Classes}} &\multicolumn{6}{c|}{\textcolor{black}{\bf KS probability}}\tabularnewline

\hline
&&&&&&\\
$z<0.15$&\multicolumn{1}{l}{Log($L_{1.4GHz}$)}&
$Log(L_{\it OIII}$)
& Log(BH) 
&Log($L_{acc}/L_{Edd}$) 
&\multicolumn{1}{l} {Log($M _{\ast}$)} 
&\multicolumn{1}{l|} {D$_{n}4000$}\\
 \hline

\textcolor{black}{FRI vs FRII-LERG} & \textcolor{black}{\boldmath $<10^{-3}$} & 0.18 &$3.7\times10^{-3}$&$3.5\times10^{-3}$ &$5.7\times10^{-2}$&0.06\\
\textcolor{black}{FRI vs FRII-HERG} &\textcolor{black}{\boldmath $<10^{-3}$}  & \textcolor{black}{\boldmath $<10^{-3}$} &\textcolor{black}{\boldmath $<10^{-3}$}&\textcolor{black}{\boldmath $<10^{-3}$}&\textcolor{black}{\boldmath $<10^{-3}$}&\textcolor{black}{\boldmath $<10^{-3}$}\\
\textcolor{black}{FRII-LERG vs FRII-HERG} &$1.9\times10^{-2}$ &\textcolor{black}{\boldmath $<10^{-3}$ } & \textcolor{black}{\boldmath $<10^{-3}$}  & \textcolor{black} {\boldmath $<10^{-3}$ } &$3.2\times10^{-3}$&\textcolor{black}{\boldmath $<10^{-3}$}\\

\hline
&&&&&&\\
\textcolor{blue}{$z<0.05$ }
&\multicolumn{1}{l}{Log($L_{1.4 GHz}$})& Log($L_{\it OIII}$)  &Log(BH) &Log($L_{acc}/L_{Edd}$) &\multicolumn{1}{l} {Log($M _{\ast}$)} &\multicolumn{1}{l|} {D$_n$4000}\\
\hline
\textcolor{blue}{FR0 vs small FRI} & --               &$8\times10^{-3}$ &$3.2\times10^{-2}$ &\textcolor{black}{\boldmath$<10^{-3}$} &0.33     &0.33 \\
\textcolor{blue}{FR0 vs FRI}& --               & 0.34            &$2.1\times10^{-3}$ &0.20                &0.01      & 0.37\\
\textcolor{blue}{small FRI vs FRI}                   & 3.8$<10^{-3}$ &0.19&$5.7\times10^{-2}$& 0.42             &0.06      &0.84\\
\hline
\end{tabular}
\label{ks}
\end{table*}

\begin{figure*}
\centering  

\plottwo{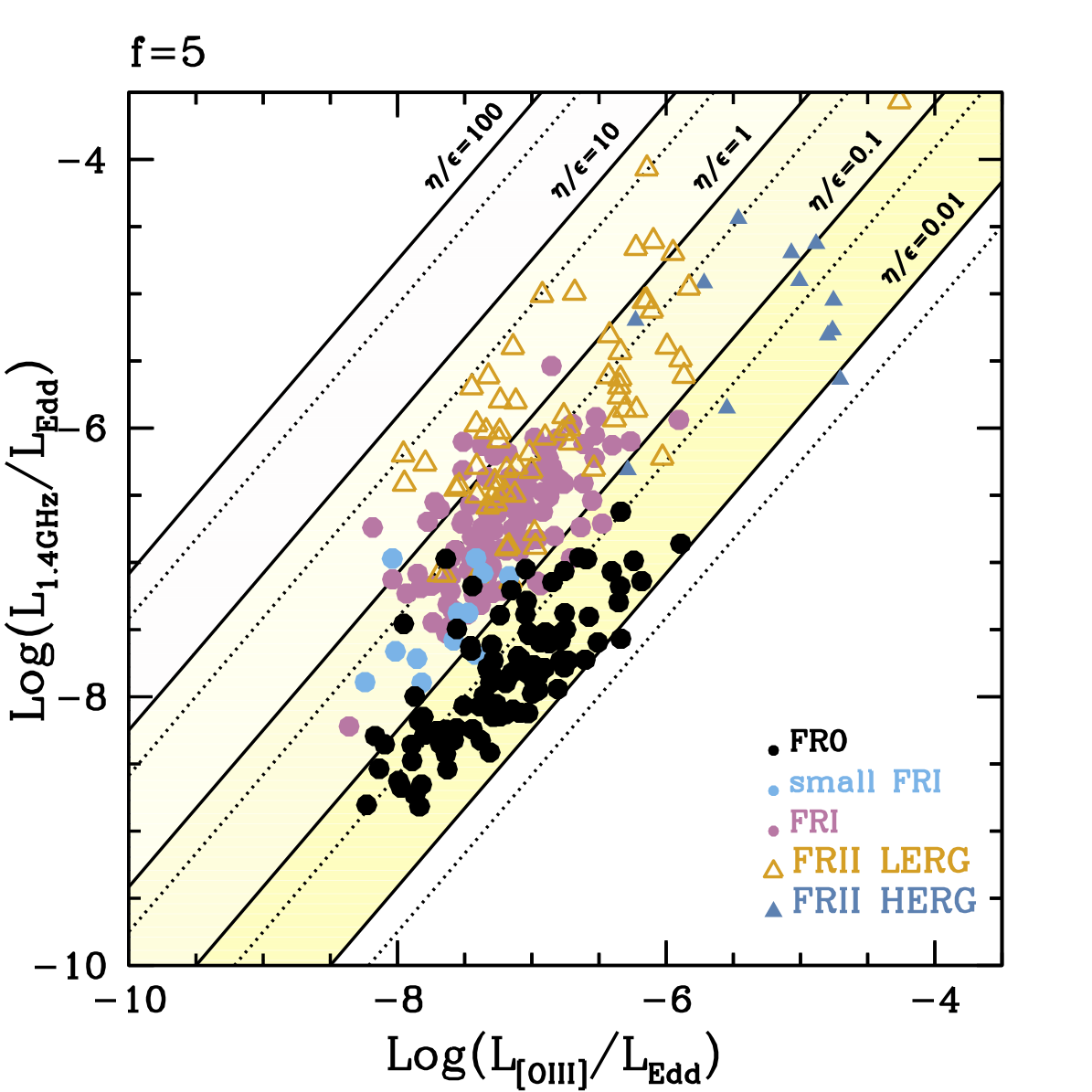}{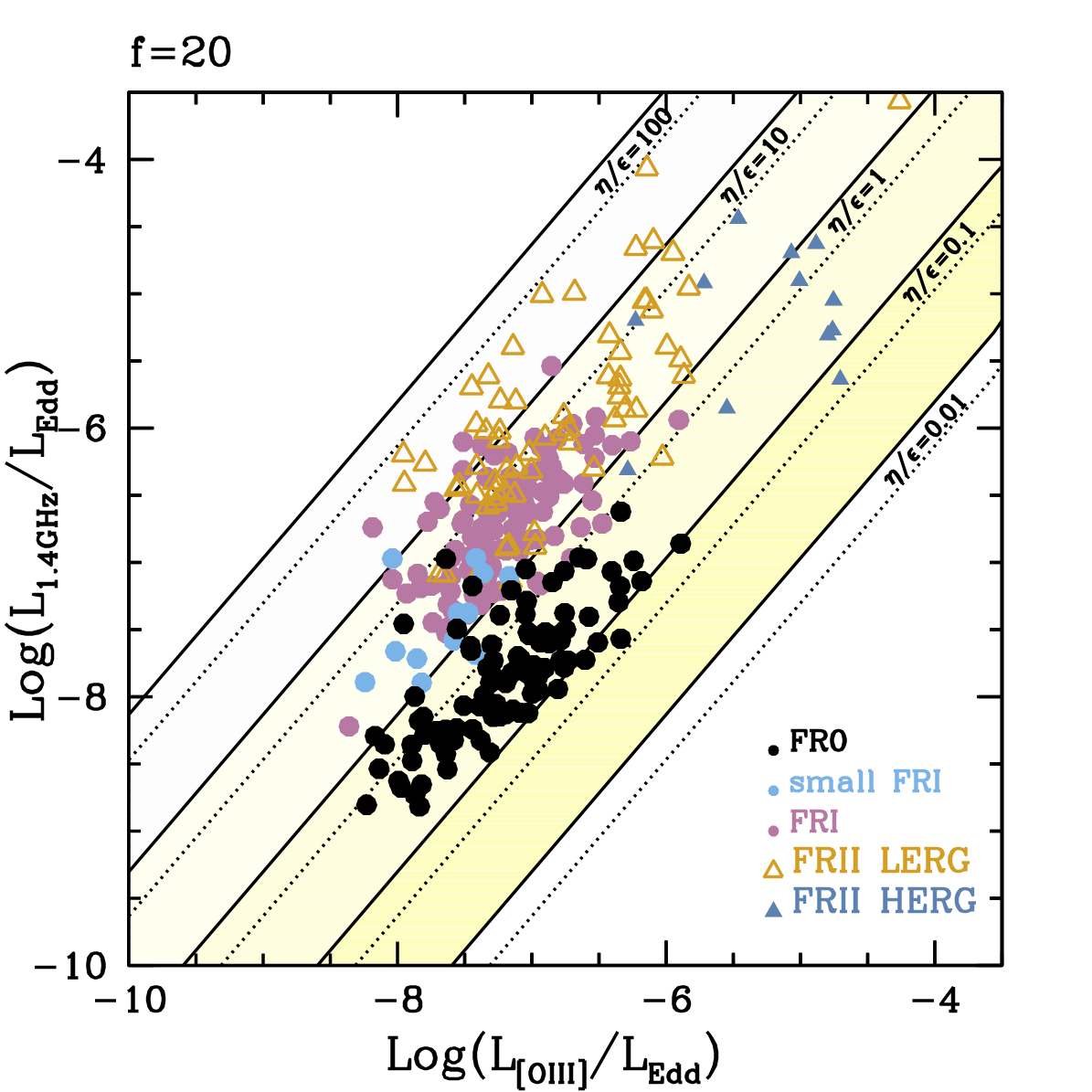}
\caption{$L_{1.4GHz}/L_{Edd}$ versus $L_{[OIII]}/L_{Edd}$ of FRcat sources compared to the predicted values estimated by equation (\ref{eq:2})  assuming $f=5$ ({\it left panel}) and $f=20$ ({\it right panel}). Each line in the plots corresponds to a different value of $\eta/\epsilon$. Solid and dotted lines corresponds to $M_{BH}=10^{7.5}$ M$_{\odot}$ and $M_{BH}=10^{9.5}$ M$_{\odot}$, respectively.}
\label{eff2}
\end{figure*}

\section{Jet power versus accretion power}
In this section we explore the jet-accretion link in mJy radio galaxies to find possible intrinsic differences in their nuclear engine.
Following \cite{Shankar}, we define the jet and the radiation power as: $P_{jet}=\eta \dot{M} c^2$ and $L_{acc}=\epsilon \dot{M} c^2$, being $\eta$ and $\epsilon$, the fraction of gravitational energy converted into jet power and thermal radiation, respectively. Combining the two relations we obtain:
\begin{equation}\label{eq:1}
P_{jet}= (\eta/\epsilon) L_{acc}
\end{equation}
\noindent 
The (\g) ratio directly measures the ability of the system to channel gravitational energy into the jet rather than to dissipate it in thermal radiation.

The radiative  power  can be related to the [OIII] luminosity through: $L_{acc}=L_{OIII}\times 3500$ (see Section 2), while the kinetic power, expressed as a function of the radio luminosity ($L_{rad}$), is generally written as: $P_{jet}=K L_{rad}^{\Gamma}$.

A jet power-radio luminosity relation was at first proposed by \cite{willott99}:

\begin{equation}\label{eq:2}
 P_{jet}=4\times10^{35}f^{3/2}L_{1.4}^{0.86}
\end{equation}

Here the original relation, that uses the radio luminosity at 151~MHz, is re-scaled to 1.4~GHz adopting a radio spectral index $\alpha$=0.8  \citep{heckman14}. The luminosity is in units of $10^{25}$~Watt~Hz$^{-1}$.\\ 
As a starting point, \cite{willott99} assumed that the jet energy is mainly stored in the lobe and/or utilized to expand the radio source and considered the radiative losses negligible.
They provided a minimal estimation of the internal energy in an equipartition regime (i.e. the internal energy is almost equally distributed between magnetic field and relativistic particles) and then divided this quantity by the source age. The $f$ factor (included in the normalization) absorbs all the uncertainties on the physical state of the lobes, such as the particle composition and their spectral distribution, volume filling factor, possible deviation from the equipartition condition, presence of internal turbulence, and fraction of internal energy lost as work done during the lobe expansion (assumed to be $\sim 50\%$ of the internal energy). Among these, the number of protons per electron present in the relativistic plasma is the most relevant one.
The analysis of \cite{willott99}, based on FRII and Steep Spectrum Radio Quasars, constrained $f$ between 1 (light jet) and 20 (heavy jet).
Later, \cite{daly2012}, investigating  31 FRII radio galaxies with an accurate radio characterization, found a $P_{jet}$ relation in substantial agreement with equation (\ref{eq:2}) and a value of $f \sim 5$.

The extension of these studies to sources in gas-rich environments (mainly FRIs) was viable after the launch of the Chandra satellite. The discovery of X-ray cavities around radio lobes suggested a different approach to estimate P$_{\rm jet}$. The jet power
could be deduced considering the energy spent by lobes to displace the surrounding gas and the age of the cavity (see \cite{birzen08} for details on the age calculation). 
The required energy (i.e. the enthalpy H) to excavate the medium is the sum of the work done by the lobes (pV) and their thermal energy.  
The enthalpy is assumed to be H=4pV if the lobe is dominated by relativistic particles \citep{McNamara2007}.
\cite{cavagnolo10} studied 16 giant radio galaxies (mainly FRIs) and found $P_{jet} \propto L_{1.4}^{0.75(\pm0.14)}$, in quite agreement with the Willott's relation.  

Finally, we mention a study of 15 radio galaxies \citep{MH} where a  relation between the jet power, estimated from X-ray cavities, and the radio core luminosity was explored. The authors reported a relation, $P_{jet} \propto L_{core}^{0.8}$, similar to those deduced by  \cite{willott99} and \cite{cavagnolo10} allowing to extend the study of jet kinetic power to also small and unresolved \rgs.  The assumption here is that beaming Doppler boosting effects do not amplify the 1.4 GHz radiation. We are quite confident that this condition is satisfied by our \rgs, considering that suspected BL LAC objects have been excluded by the FR0 cleaned sample. Incidentally, we note that the core contribution does not significantly affect the extended \rgs. As a check, we obtained a rough estimation of the  "extended" 1.4 GHz luminosity subtracting the FIRST peak flux (assumed to be a proxy of the core emission) from the total NVSS flux. This test was possible for more than 90\% of FRI and FRII sources.
The luminosities generally changed less than 0.2 dex on logarithmic scale with no significant impact on our study (see next section).\\

Considering the above-mentioned caveats, we will handle the $f$ parameter as an unknown variable in the following. Moreover,  when equation ({\ref{eq:2}) is exploited to estimate the 
\g~ ratios of FR0s and small FRIs, their  NVSS luminosities will be treated as upper limits.

\subsection{Predicted luminosities in systems with different accretion efficiency ratios}

Relation (\ref{eq:2}) can be re-written in order to have the 1.4~GHz luminosity as subject: for an [OIII] luminosity as input, it then allows to estimate the expected radio luminosity for any value of $f$ and \g. 

The {\bf predicted} $L_{\rm 1.4GHz}$ values, obtained assuming $L_{\rm [OIII]}$ ranging from 39 to 45, are shown in Figure \ref{eff2} for $f=5$ ({\it left panel}) and $f$=20 ({\it right panel}). The luminosities are rescaled to the Eddington luminosity (L$_{Edd}$), using two  different black hole masses ($M_{BH}=10^{7.5}$ M$_{\odot}$ and $M_{BH}=10^{9.5}$ M$_{\odot}$) to cover the mass range observed in the FRcat sources.
Each line in the plots corresponds to a different value of $\eta/\epsilon$. Solid and dotted lines correspond to 
$M_{BH}=10^{7.5}$ M$_{\odot}$ and $M_{BH}=10^{9.5}$ M$_{\odot}$, respectively.
Note that a change of the black hole mass has not an important impact on the predicted $\eta/\epsilon$ curves. 
As stressed in the previous section,  the FRI and FRII radio luminosities could be overestimated by $\sim 0.2$ dex.  Considering the intrinsic spread of each class in each plot, this effect is negligible. For small and compact sources the \g~ values should be considered upper limits, being unconstrained the contribution of the radio core to the total L$_{1.4GHz}$ luminosity.
Comparing the two panels, it appears also evident that a variation of  $f$   only translates the $\eta/\epsilon$ curves, preserving the relative position of the different classes.\\
As expected, LERGs and HERGs, having different accretion rates (i.e. different L$_{\rm [OIII]}/L_{\rm Edd}$), occupy different parts of the plot. FR0s, being compact radio sources by definition,  populate the lower left corner. 
However, a more careful inspection of Figure \ref{eff2} shows that
FRIs and HERGs preferentially fall in different $\eta/\epsilon$ strips and that LERGs are spread along the y axis.
It seems that jet-disk systems in HERGs favour a thermal dissipation of the gravitational power, while jets of different powers can be launched by very similar inefficient accretion flows. However, the implicit assumption here is that the normalization (i.e. $f$) of equation (\ref{eq:2}) is the same for each FR classes.

\subsection[Exploring]{Eploring the [\g-f] parameter space of the FRcat sources}
In order to better investigate the problem, we then decided to explore the [\g-f] parameter space of each class. 
This time, the $\eta/\epsilon$ values  were determined via equation (\ref{eq:2}) utilizing the {\bf observed}} average  $L_{\rm [OIII]}$, $L_{\rm 1.4GHz}$ and $L_{\rm Edd}$ luminosities in Table \ref{stat} and running $f$  from 1 to 20.
The (\g-f) pairs that do not satisfy the condition  $L_{acc}\le L_{\rm Edd}$ were excluded.  \\

\begin{figure}[!ht]
\centering  
\plottwo{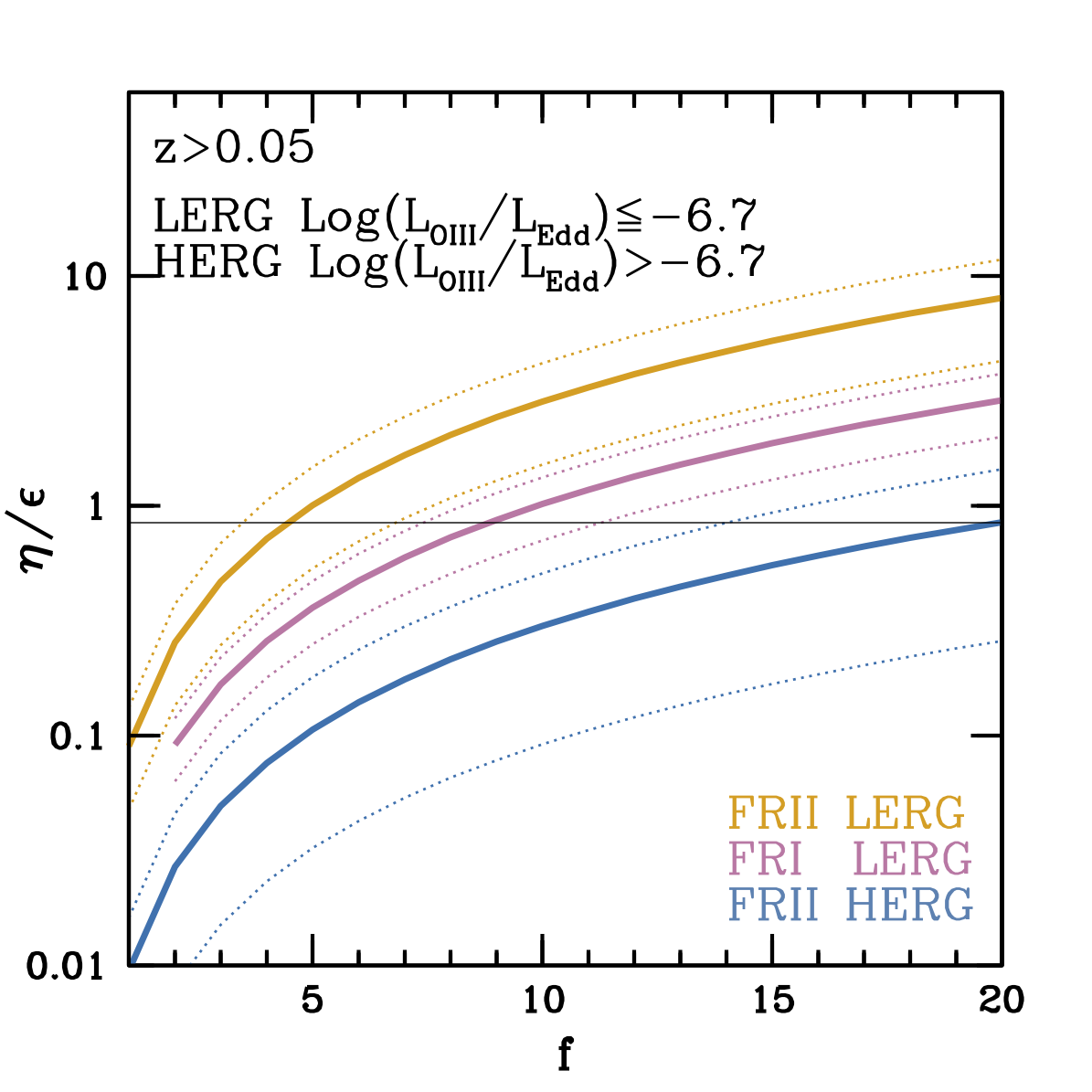}{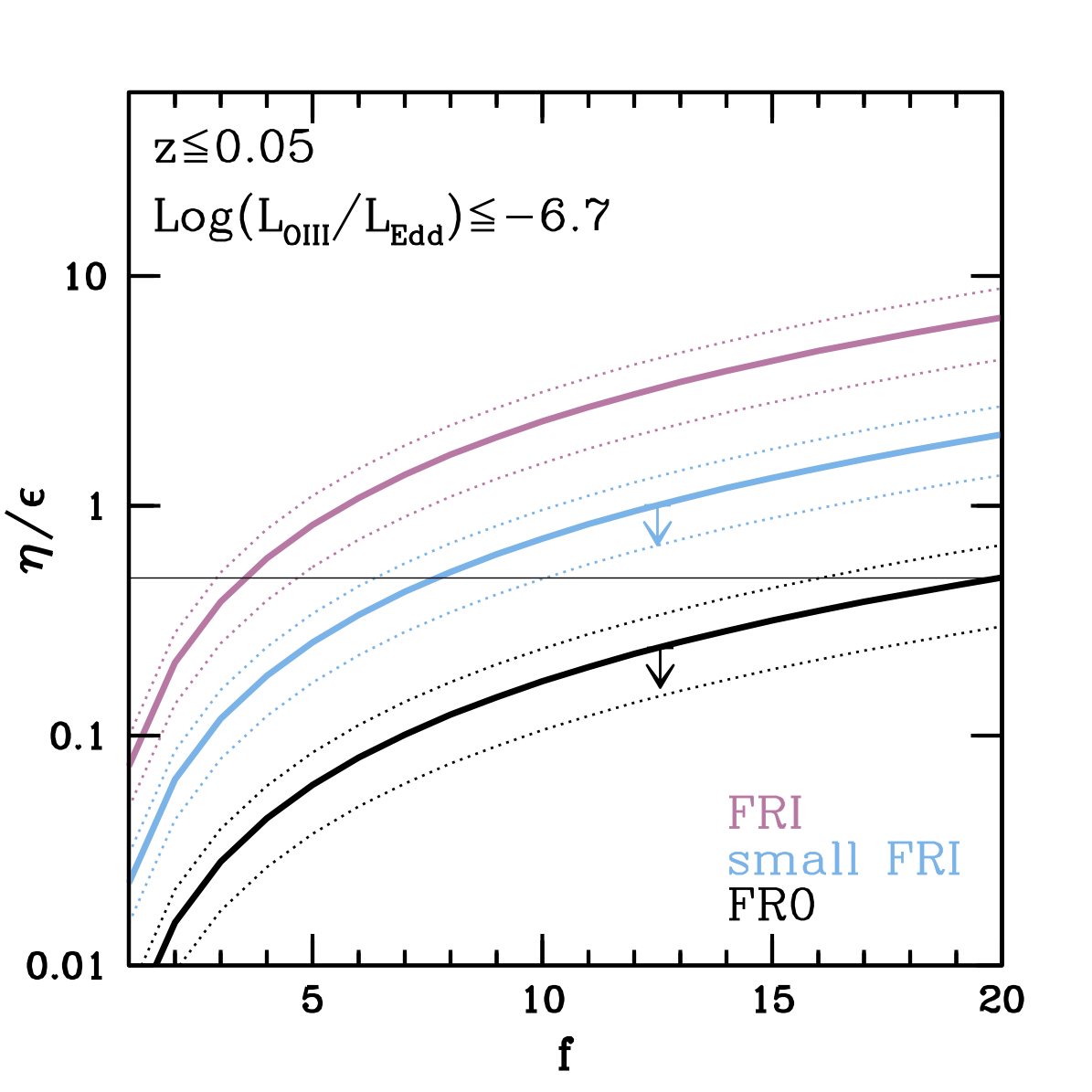}
\caption{{\it Left Panel} -- [$\eta/\epsilon$-$f$] parameter space for FRIs, FRII-LERGs and FRII-HERGs with z$>0.05$. The curves 
are obtained via equation (\ref{eq:2}) assuming the average $L_{\rm [OIII]}$ and $L_{\rm NVSS}$ luminosities (and standard deviation $\sigma$)  in Table \ref{stat}. Dotted lines correspond to 1-$\sigma$ uncertainty. The separation at $Log(L_{\rm [OIII]}/L_{\rm Edd})=-6.7$ between \hgs~ and \lgs~ is based on Figure \ref{eff2}. The intersections of the horizontal line with the curves mark the $f$ values required  to have a same efficiency ratio ($\eta/\epsilon=0.98$ in this case) in each class. {\it Right Panel} -- Same plot for  FR0s, small FRIs and FRIs with z$\le0.05$. The horizontal line corresponds  to $\eta/\epsilon=0.49$. Vertical arrows indicate that the \g~ values traced by the curves have to be intended as upper limits.}

\label{top}
\end{figure}

In Figure \ref{top}-({\it left panel}), the $f$ and $\eta/\epsilon$ permitted values for FRII  and FRI  radio galaxies at $z>0.05$ are shown for two different accretion rates. The separation at $Log(L_{\rm [OIII]}/L_{\rm Edd})=-6.7$ is based on Figure \ref{eff2}. 
As already noted,  the efficiency ratio increases from FRII-HERGs to FRII-LERGs if $f$ is kept constant.  Different classes could have the same \g~ ratio only if the normalization of equation (\ref{eq:2}), i.e. $f$, is allowed to vary.

If the main source of uncertainty included in $f$ is the plasma particle content \citep{willott99}, the condition of equal $\eta/\epsilon$ could be reached in FRIs and FRII-HERGs only assuming that jets are lighter in the former sources.
Although not completely rejectable (our understanding of the particle acceleration near the black hole is really poor), this hypothesis does not seem to be supported by the observations of radio structures on kpc scales.
The decelerated and less collimated jets seen in FRIs are indeed suggestive of strong interactions with the environment and mass loading through mixing in turbulent layers \citep{perucho}. 
X-ray studies of radio lobes and gaseous environments of FRIs and FRIIs \citep{ineson,cro2018} indicate indeed that core-brightened Radio Galaxies contain more protons than edge-brightened Radio Galaxies.

Another source of uncertainty, that could be invoked to satisfy the equal \g~ condition, is the ambient medium \citep{cavagnolo10}.
Jets that propagate in a dense environment have to spend  more internal energy pushing away the surrounding gas.
A larger corrective factor (thus a larger $f$) should then be included in the normalization of Willott's relation (\ref{eq:2}) if a radio source lives in a rich environment. 
Again, this conflicts with the observations.  FRIs (that should have a smaller $f$ than HERGs for equal \g~)  are preferentially found in groups or clusters when bright (Jy) radio galaxies are considered \citep{gendre}. Moreover, 
no environmental difference  between FRIs and FRIIs is observed in the Local mJy Universe \citep{massaro}. \\
In summary, it seems unlikely that \rgs, powered by different accretions, choose the same dissipative channel. It is more plausible that FRIs convert  most of their gravitational energy into jets power and FRII-HERGs into thermal radiation.\\

FRII-LERGs represent a more complex class. They have radio morphologies and particle content \citep{ineson} similar to FRII-HERGs but habit older galaxies, have more massive black holes and a smaller accretion rate. 
In addition, they have the largest $\eta/\epsilon$ ratios (Fig. \ref{eff2}) despite their marked similarity with FRIs.  
As proposed by \cite{duccio}, a possibility is that FRII-LERGs are old  HERGs that, exhausted their fuel, have changed the accretion mode. In this case, the values of \g~  are meaningless, as Equation (\ref{eq:2}) cannot be applied anymore being the nuclear region and the extended radio structures temporarily disconnected. On the other hand, theoretical studies show that inefficient accretion configurations between an ADAF and a cold efficient disk can exist (see Figure  1 of \cite{yuan}). If, for some reasons more viscous dissipated energy is transferred into electrons and radiated away, the ADAF accretion flow can enter into a more luminous hot accretion flow regime  \citep{xie}. If the  electron  cooling  becomes  too strong, the matter in accretion collapses in a cold disk or in cold dense clumps embedded in a hot flow \citep{yuan}. We could then be observing the inverse trend.

Another option is that FRII-LERGs are a separate class and not a product of the FRII-HERG evolution.
The observed $L_{OIII}/L_{Edd}$ spread of FRII-LERGs (see Fig. \ref{eff2}) could then be simply due to the co-existence of ADAF configurations with different electron cooling. The high $\eta/\epsilon$ ratios are however difficult to explain, unless, for example,
more extreme conditions of the black hole properties are assumed for this class (see discussion below).

The [\g-f] parameter space of \rgs at z$\le$ 0.05 (i.e FR0, small FRIs and local FRIs) is shown in Fig. \ref{top}-({\it right panel}). 

The plot is similar to that observed for sources at higher redshift: the  efficiency ratio increases from FR0s to extended FRIs if $f$ is kept fixed.  Also in this case equal  $\eta/\epsilon$ values  would require $f$ changing from a class to another one, implying possible
different intrinsic (jet content)  or external (environment) conditions.
FR0s, that are less able to channel energy into the jets (\g~is always  less than 1) should expel heavier jets or be embedded  in a very dense environment. The first hypothesis is difficult to test (in particular if the radio emission is not extended on large scales). The second one is more intriguing.
The idea that a hostile ambient inhibits  the jet expansion of small/compact \rgs  is indeed plausible. However, the observations do not support this view.
A study based on the galaxy richness  around the FRCat sources at z$\le 0.05$ does not reveal any FR0 over-density \citep{capettienv}. In addition, an X-ray study of the galaxy cluster Abell~795 having a FR0 at its center (Ubertosi et al 2021 A$\&$A submitted, Ubertosi Master Thesis\footnote{https://amslaurea.unibo.it/21460/})  found  gas density and temperature typical 
of clusters hosting more extended central radio galaxies.\\
Figures \ref{eff2} and \ref{top} are suggestive of another viable interpretation.
The different radio luminosities observed in LERGs having comparable  accretion rates ($L_{\rm [OIII]}/L_{\rm Edd}$) might indicate that similar nuclear engines impart diverse  accelerations to the outflowing plasma. 
One of the most accredited model for the jet production \citep{BZ}
links  the jet kinetic power to the properties of the black hole, i.e.  mass, spin (a) and  magnetic field at its horizon ($\Phi$): $P_{BZ} \propto \Phi^2 a^2 M_{BH}^2$. Since LERGs have similar accretion rates
and black hole masses (Table \ref{stat}), the vertical  displacement of the FRcat [$\eta/\epsilon$-$f$] curves could directly map different values of $a$ and/or $\Phi$. In this view, FR0s (at least those with $Log(L_{\rm [OIII]}/L_{\rm Edd})<-6.7)$ should have extremely slow black holes and/or weak magnetic fluxes, while  FRII-LERGs, assumed not evolved HERGs, the largest values of $a$ and/or $\Phi$.\\

Finally, we noted that Equation (\ref{eq:2}) does not take into account mildly relativistic winds that could contribute to the total kinetic budget ($P_{tot}=P_{jet}+P_{winds}$). 

An Advection Dominated Inflow-Outflow Solution (ADIOS) proposed by \cite{BB99} predicts the presence of matter outflows that, exceeding the amount of material crossing the black hole horizon, favours a low accretion rate. Magneto-hydrodynamic (MHD) simulations of ADAF \citep{sadowski} also predict winds. These are expected to be  less energetic than jets 
unless the  black hole spin and/or the magnetic flux are small\footnote{FR0s could be in this condition and dissipate more of their gravitational power into winds.}. Moreover,  \cite{liska} showed that both jets and magnetically driven winds can be produced by AGNs with a thin accretion disk (and a fast spinning black hole). 

On the observational side, several works attest to the existence of outflows in bright radio galaxies.
A very recent work by \cite{bia}, exploring the innermost jet profile of several radio-loud AGN using VLBI data, has confirmed the
existence of thick disk-launched layers surrounding the HERG jets and of less prominent sheaths, anchored to the innermost accretion regions, in LERGs.
X-ray studies also confirm the co-existence of jets and outflows in the nuclear region of powerful radio galaxies \citep{torresi,tombesi,costantini}. The velocity of these winds covers a wide range of possible values, in some cases, it can be as fast as  $\sim 0.2c$. The measures  of very fast outflows with velocities reaching an appreciable fraction of \emph{c} are technically difficult and probably limited by the transiency of the event. However, if consolidated, they will attest to the important role of the winds in the energy balance.

\section{Conclusions}
In this paper the study of the mJy sources of the FRCat catalogs was performed following two different approaches.
At first, we performed a statistical analysis of the main observables and compared the average properties of the different classes. Then we explored the jet-accretion system exploiting  the known relations that link $L_{\rm [OIII]}$ and L$_{\rm 1.4GHz}$ to the accretion (thermal) and jet kinetic power, respectively.

The main results of our statistical analysis are summarized below:
\begin{itemize}
    \item FRIs compared to FRII-HERGs show more massive black holes, smaller accretion rates (expressed in terms of $L_{\rm acc}/L_{\rm Edd}$), larger stellar masses, and a more evolved stellar population;
    \item FRII-LERGs are more similar to FRIs than FRII-HERGs;
    \item No significat difference is observed between small FRIs and FRIs at z$<0.05$. All the local sources are hosted in massive galaxies with no recent star forming activity and  have comparable  low accretion rates and black hole masses;
    \item FR0s show $M_{\star}$ and D$_n$4000 typical of evolved system. Their Log($L_{\rm acc}/L_{\rm Edd}$) ratio extend to values higher than local ($z\le 0.05$)  FRIs.
\end{itemize}

These results suggest that, in the mJy Universe, the majority of radio galaxies within $z\le0.15$ are in a late stage of their life.
The only exception is represented by the FRII-HERG class which is however poorly populated.

From a comparison between Jy and mJy FRII-LERGs, it emerges that lower radio flux density sources have, on average,  FRI-like characteristics, while  FRII-LERGs of the 3C sample  are more 'active'  with intermediate properties between FRIs and FRII-HERGs \citep{duccio}. 
This points towards an evolutionary scenario in which  FRII-LERGs are aged FRII-HERGs. Once  the nuclear cold fuel has been consumed, the accretion configuration becomes hot and inefficient while the extended radio structures conserve traces of the past activity. 
 It has been shown that a  wide range of configurations between thick hot flow and thin cold disk is stable. If, for example, a strong and turbulent magnetic field permeates the accreting matter, MHD instabilities/magnetic reconnections can further heat the electrons that can radiate away giving origin to more luminous hot accretion flows.
Similar accretion configurations could account for the wide range of Log($L_{\rm [OIII]}/L_{\rm Edd}$) in Fig. \ref{eff2},  and even more for the 
higher [OIII] luminosities observed in the 3C FRII-LERGs.

We cannot however reject the hypothesis that FRII-LERGs are a separate and independent class  with an inefficient accretion regime able to produce extended FRII radio structures. This  breaks the correspondence between efficient/inefficient accretion and strong/weak jets, making appealing other options directly involving the black hole spin and /or the magnetic field at its  horizon.

To further investigate this possibility we focused on the efficiency ratio parameter ($\eta/\epsilon$), that quantifies 
the capability of a source to convert gravitational energy into jet power rather than in thermal radiation. We exploited the 
relations $P_{jet}=K(f)L_{radio}^{0.86}$ and L$_{\rm acc}=3500 \times L_{\rm [OIII]}$ that, although empirical and 
affected by several uncertainties (absorbed by the $f$ parameter) allow to directly  relate jet kinetic and accretion powers to observed luminosities.
Aware of the intrinsic limitation of this approach, we compared the \g~  ratios of the different classes considering two  main sources of uncertainties: the particle composition of the relativistic plasma \citep{willott99} and the work done by the jets on the surrounding medium \citep{cavagnolo10}. \\
We observe that:
\begin{itemize}
\setlength\itemsep{0em}
    \item FRIs and FRII-HERGs have different efficiency ratios. A similar $\eta/\epsilon$ in the two classes would require
    jet compositions and environment conditions not supported by the observations. In FRIs  the gravitational energy is preferentially channeled into the jets, in FRII-HERGs mainly dissipated by thermal photons.
    Although our study does not include sub-relativistic/mildly relativistic matter outflows, winds probably contribute to the total energy budget. In FRIs, jets launched  by the Blandford \& Znajek \citep{BZ} process should co-exist with winds produced by the ADAF itself. Outflows of matter are indeed theoretically predicted in the inefficient accretion regimes (see the ADIOS model) and are also revealed in  MHD simulations. In FRII-HERGs both the Blandford \& Znajek \citep{BZ} and the Blanford \& Payne  \citep{BP} mechanisms could then be  at work. The former launches jets extracting  energy by the spinning black hole, the latter produces centrifugally driving outflows of matter from a magnetized disc.
    The recent VLBI study of inner jet profiles of radio galaxies \citep{bia} strongly supports this scenario.
    
    \item  The wide range of Log($L_{\rm 1.4GHz}/L_{\rm Edd}$) observed in radio galaxies with similar Eddington normalized [OIII] luminosities ($L_{\rm [OIII]}$/L$_{\rm Edd}\le -6.7$) might indicate that neither the black hole mass nor the rate of the mass accretion are the key parameters to explain the class segregation of LERGs.
    If the difference originates in the nuclear engine, then the spin of the black hole and/or the magnetic field threading its horizon are  fundamental ingredients. Following  \cite{BZ} the jet propulsion could be less potent in FR0s than in FRIs,  because the black hole spins are slower and/or the magnetic field is weaker. Extending this interpretation to radio sources at z$> 0.05$, the high Eddington normalized radio luminosities of FRII-LERGs (assumed as a class on its own) would imply black holes with the fastest spin and/or most  intense magnetic field. 
    \item  Assuming typical values of $\epsilon \sim0.1$ for efficient disks  and $\epsilon \sim$ 10$^{-2}$-$10^{-3}$ for ADAFs,  
    an average  $\eta$ via equation (\ref{eq:2}) can be derived for each class. 
    In mJy sources the fraction of gravitational power conveyed by the jets is modest, at most 10$\%$ in HERGs (excluding winds) and a few percentages in LERGs (despite their larger $\eta/\epsilon$ ratios).
    FR0s are the more extreme case, with  $\eta <$ a few $10^{-3}$. 
\end{itemize}

\section*{Acknowledgments}
We thank Andrea Merloni for  useful discussions and the anonymous referee for his/her constructive comments.
ET acknowledges the financial contribution from the agreement ASI-INAF n. 2017-14-H.O. This research made use of the NASA/ IPAC Infrared Science Archive and
Extragalactic Database (NED), which are operated by the Jet Propulsion Laboratory, California Institute of Technology, under contract with the
National Aeronautics and Space Administration. Part of this work is based on archival data,  software online services provided by the ASI Space Science Data Center (SSDC).

\bibliography{FR_r}{}
\bibliographystyle{aasjournal}



\end{document}